\documentclass[10pt, conference]{IEEEtran}
\IEEEoverridecommandlockouts
\usepackage{amsmath,amsfonts}
\usepackage{array}
\usepackage[caption=false,font=normalsize,labelfont=sf,textfont=sf]{subfig}
\usepackage{textcomp}
\usepackage{stfloats}
\usepackage{url}
\usepackage{verbatim}
\usepackage{graphicx}
\usepackage{bbm}
\usepackage{dsfont}
\usepackage{caption}

\def\BibTeX{{\rm B\kern-.05em{\sc i\kern-.025em b}\kern-.08em
    T\kern-.1667em\lower.7ex\hbox{E}\kern-.125emX}}
\usepackage{balance}

\usepackage{amsmath,amsfonts}
\usepackage{array}
\usepackage[caption=false,font=normalsize,labelfont=sf,textfont=sf]{subfig}
\usepackage{textcomp}
\usepackage{stfloats}
\usepackage{url}
\usepackage{verbatim}
\usepackage{graphicx}
\usepackage{cite}
\usepackage[ruled,vlined,linesnumbered]{algorithm2e}
\usepackage{algpseudocode}
\usepackage{xcolor}
\usepackage{multirow}
\usepackage{amsfonts}
\usepackage{xspace}
\usepackage{colortbl}
\usepackage{makecell}
\usepackage{microtype}
\usepackage{enumitem}
\usepackage{pifont}
\usepackage{booktabs}
\usepackage{amsthm, amssymb}
\usepackage[switch]{lineno}
\usepackage[hidelinks]{hyperref}
\usepackage{tikz}
\usetikzlibrary{positioning}
\usepackage{graphicx}
\usepackage{tikz}
\usepackage{graphicx}
\usepackage{tabularx}
\usepackage{xurl}
\usepackage[most]{tcolorbox}
\usetikzlibrary{positioning, arrows.meta, shapes.geometric}
\newtheorem{definition}{Definition}

\definecolor{citeblue}{RGB}{0,102,180}

\hypersetup{
    colorlinks=true,
    linkcolor=black,
    citecolor=citeblue,
    urlcolor=citeblue
}

\newcommand{\tool}{\textit{EvoEye}\xspace}
\newcommand{\monitor}{\textit{FusionMonitor}\xspace}
\newcommand{\evo}{\textit{BlindSpotEvolver}\xspace}

\begin{document}


\title{\tool{}: Simulation-Guided Self-Evolving Safety Monitoring for Autonomous Driving Systems}

\author{
    \IEEEauthorblockN{Mingfei Cheng$^{\dagger,1}$, Lionel Briand$^{2,3}$, and Xiaofei Xie$^{1}$}
    \IEEEauthorblockA{$^1$Singapore Management University, Singapore $^2$University of Ottawa, Canada}
    \IEEEauthorblockA{$^3$Lero Centre, University of Limerick, Ireland}
    \vspace{-2.1em} 
}


\maketitle

\begingroup
\renewcommand{\thefootnote}{\textdagger}
\footnotetext{This work was done while the author was visiting the University of Ottawa.}
\endgroup

\begin{abstract}
Safety monitoring is essential for autonomous driving systems (ADSs) to detect unsafe behaviors and provide timely warnings before safety failures occur. However, ADSs are complex systems comprising multiple interacting modules that produce heterogeneous internal signals with complex temporal dependencies. These signals primarily reflect local module states, making it difficult to predict system-level safety failures such as collisions.

To address these challenges, we propose \tool{}, a simulation-guided framework for self-evolving safety monitoring of ADSs. Specifically, we design \monitor{}, a learning-based monitor that models internal module signals and their temporal interactions to predict system-level safety outcomes. We further develop \evo{}, an evolutionary algorithm that guides simulation to explore scenarios where the monitor produces incorrect predictions. We then use these scenarios to iteratively retrain the monitor, enabling it to improve its predictions in challenging contexts. 
We evaluate \tool{} using Baidu Apollo and CARLA across representative highway and urban scenarios. At a target frame-level false positive rate (FPR) of $0.05$, \monitor{} improves Recall over the strongest baseline by $15.2$ and $23.8$ percentage points in the two scenarios, with a CPU inference latency of only $6.26$~ms. Under the same evolution budget, \evo{} improves F1 by up to $9.7$ percentage points over the best baseline on previously missed unsafe scenarios.
\end{abstract}

\section{Introduction}

Autonomous vehicles (AVs) have advanced rapidly in recent years and can improve transportation safety, accessibility, and efficiency~\cite{paden2016survey}. 
Autonomous driving systems (ADSs) serve as the core intelligence of AVs, enabling them to perform dynamic driving tasks. In a typical modular architecture, the ADS software stack processes data from sensors such as cameras, LiDAR, and radar to perceive the surrounding environment. The resulting information is then used for behavior prediction, decision-making, motion planning, and vehicle control~\cite{badue2021selfdriving}.
As safety-critical systems, ADSs must operate safely in complex and dynamic environments, where unexpected failures can lead to severe consequences. Runtime safety monitoring is therefore essential for promptly detecting anomalous behaviors and identifying potential safety hazards during operation~\cite{watanabe2018runtime}.

Existing runtime safety monitoring approaches for ADSs mainly follow two directions. One line of work assesses driving safety using predefined physical constraints and formal safety specifications~\cite{mehdipour2023formal}. Representative approaches employ Time-to-Collision (TTC)~\cite{ttc}, Responsibility-Sensitive Safety (RSS)~\cite{shalev2017rss}, and reachability analysis~\cite{althoff2014online} to evaluate collision risks based on vehicle states and trajectories. Other approaches, such as REDriver~\cite{sun2024redriver}, use signal temporal logic specifications to monitor planned trajectories for potential safety violations. Manually specifying safety properties that capture failures arising from complex interactions among ADS modules remains difficult. 
Another line of work focuses on monitoring specific driving models to predict potential safety failures. These approaches analyze model-related information, such as input reconstruction errors~\cite{stocco2020misbehaviour}, attention maps~\cite{stocco2022thirdeye}, and trajectory prediction errors~\cite{farid2023task}. However, their focus on specific models or failure mechanisms may limit their ability to accurately predict system-level safety failures, particularly when relevant warning signals span multiple ADS modules.
More fundamentally, both lines of work rely on manually specified safety properties or training datasets that may not adequately capture system-level failure patterns. Recent studies~\cite{stocco2022confidence,torfah2022learning} have attempted to improve monitors using prediction errors observed during executions obtained through random sampling. However, these strategies may not consistently generate informative executions, limiting the effectiveness of monitor improvement.

To bridge this gap, we investigate self-evolving runtime safety monitoring for ADSs, aiming to predict system-level safety failures from runtime signals across multiple internal modules and continuously improve prediction performance. However, achieving these goals presents two main challenges. 
\ding{182} Modeling complex internal runtime signals for system-level safety prediction is challenging. At each time step, signals from perception, prediction, planning, and control reflect different aspects of the system state, and no individual module provides a complete view of system-level safety risks. The safety implications of these signals depend on their relationships across modules and how the system state evolves over time, making system-level safety failures difficult to predict.
\ding{183} Automatically evolving monitors remains challenging. A monitor may raise false alarms in safe driving scenarios or fail to warn of impending safety risks. Identifying these prediction errors requires executing the ADS in diverse driving scenarios to collect runtime signals and evaluate predictions against observed safety outcomes. However, the vast scenario space makes exhaustive testing costly and impractical. Therefore, how effectively and efficiently identifying informative executions that expose prediction errors and using them to iteratively refine the monitor remains a key challenge.

Motivated by these challenges, we propose \tool{}, a simulation-guided framework for self-evolving runtime safety monitoring of ADSs. \tool{} consists of two complementary components, \monitor{}, which models runtime signals from multiple ADS modules to predict system-level safety failures, and \evo{}, which enables automated monitor validation and iterative refinement through simulation.
To tackle the modeling challenge \ding{182}, we design \monitor{}, a learning-based monitor that employs module-specific encoders to process heterogeneous runtime signals from internal modules (i.e., perception, prediction, planning, and control). 
\monitor{} integrates the resulting representations to characterize the system state at each time step and models how these states evolve over time to predict impending system-level safety failures.
To address the second challenge \ding{183}, we develop \evo{} to automate monitor validation and refinement. Specifically, \evo{} uses the monitor's prediction errors to guide simulation toward scenarios that expose its monitoring weaknesses. To prevent the evolution process from becoming overly concentrated in a few regions, \evo{} incorporates region-aware sampling to balance exploration across the whole scenario space and avoid oversampling well-explored areas. The collected runtime signals then iteratively retrain \monitor{}, whose updated predictions guide subsequent exploration. This feedback loop enables \tool{} to progressively improve its system-level safety prediction performance.

We evaluate \tool{} on Apollo~\cite{baiduapollo} with CARLA~\cite{dosovitskiy2017carla} in representative highway and urban scenarios. At a target frame-level FPR of $0.05$, \monitor{} improves Recall over the strongest baseline by $15.2$ and $23.8$ percentage points on S1 and S2, respectively, while requiring only $6.26$~ms of CPU inference time per runtime signal context. Under the same evolution budget, \evo{} achieves the highest F1 at both target FPRs, as well as the highest AUROC and AUPRC, on the test set across two scenarios, and improves F1 on the hard cases by up to $9.7$ percentage points over the best baseline.

In summary, this paper makes the following contributions: 

\begin{itemize}[leftmargin=*]

    \item To the best of our knowledge, we propose the first self-evolving runtime monitoring framework for full-stack ADSs.

    \item We design \monitor{}, which jointly models cross-module interactions and temporal dependencies in ADS runtime signals to predict collision risk.

    \item We design \evo{}, which uses monitor prediction errors and region-aware sampling to acquire informative executions and iteratively update the monitor.

    \item We evaluate \tool{} on the industrial-grade Baidu Apollo ADS with CARLA, demonstrating effective runtime monitoring and gains from self-evolution.

\end{itemize}

\section{Background}

\subsection{Autonomous Driving Systems}\label{sec: Background-ads}

ADSs are the core intelligence of AVs, enabling vehicles to perceive the environment and make driving decisions with limited or no human intervention.
Existing ADSs can generally be categorized into two major paradigms: \textit{module-based} ADSs and \textit{end-to-end (E2E)} ADSs.
\textit{Module-based ADSs}, such as Pylot~\cite{gog2021pylot}, Baidu Apollo~\cite{baiduapollo}, and Autoware~\cite{Autoware}, are widely adopted in industrial and research settings because of their interpretable, well-structured designs.
A module-based ADS typically comprises several functional components, including \textit{Localization}, \textit{Perception}, \textit{Prediction}, \textit{Planning}, and \textit{Control}.
The localization module estimates the ego vehicle's pose by fusing information from sensors such as GPS, IMU, and LiDAR.
The perception module processes multimodal sensor inputs, such as camera images, LiDAR point clouds, and radar signals, to detect and track surrounding traffic participants.
The prediction module estimates the future behaviors or trajectories of nearby actors, while the planning module generates a feasible trajectory for the ego vehicle.
Finally, the control module translates the planned trajectory into low-level actuation commands, such as steering, throttle, and braking.
In contrast, \textit{end-to-end ADSs} have recently attracted increasing attention, driven by rapid advances in deep learning.
Instead of explicitly decomposing the driving task into multiple functional modules, E2E approaches~\cite{bojarski2016end, codevilla2018end, openpilot, roach_iccv, chitta2022transfuser, wu2022trajectory} directly map sensor observations to driving actions or future trajectories through a unified model.

In this paper, we monitor ADS using observable runtime signals within the system. For example, we obtain these signals from inter-module messages, such as perception outputs, predicted trajectories, planning outputs, and control commands.

\subsection{Scenario and Runtime Signal}\label{sec:scenario-description}

Scenarios provide a fundamental abstraction for testing ADSs by specifying the driving conditions under which they are evaluated~\cite{thorn2018framework}. Given the vast ADS scenario space, existing approaches~\cite{haq2022efficient,av_fuzzer,sharifi2023identifying,cheng2023behavexplor} typically define configurable scenario parameters to systematically explore diverse driving conditions. For example, an unprotected intersection scenario can be configured with different vehicle positions, speeds, and activation times. In this work, we vary these parameters to collect diverse ADS executions and runtime signals for monitoring self-evolution. 
Formally, we denote the scenario space by $\mathcal{X}$, where each \textit{scenario} $x \in \mathcal{X}$ is a specific parameter assignment. 
Given an ADS $\mathcal{A} = \{m_{1}, \ldots, m_{K}\}$ with $K$ modules, executing scenario $x$ in simulation produces the scenario observation $\mathbf{O}(x)$ and internal runtime signals $\mathbf{M}(x)$:
\begin{equation}
    \mathbf{O}(x), \mathbf{M}(x) = f_{\mathrm{sim}}(x,\mathcal{A}),
\end{equation}
where $f_{\mathrm{sim}}$ denotes the simulation procedure. 

The \textit{scenario observation} $\mathbf{O}(x) = (\mathbf{o}_{1}, \ldots, \mathbf{o}_{T})$ records the ground-truth information provided by the simulator during the execution of scenario $x$. 
Each \textit{scene observation} $\mathbf{o}_{t}$ captures the states of traffic actors at frame $t$, including their positions, velocities, and headings. The \textit{runtime signals} capture internal information produced by the ADS during the execution. Formally, we denote the runtime signal sequence as $\mathbf{M}(x) = (\mathbf{m}_{1}, \ldots, \mathbf{m}_{T})$, where $\mathbf{m}_{t} = (\mathbf{m}_{t}^{1}, \ldots, \mathbf{m}_{t}^{K})$ contains signals from $K$ ADS modules at frame $t$. Specifically, $\mathbf{m}_{t}^{k}$ represents the runtime signals from module $m_k$, such as intermediate outputs from perception, prediction, planning, and control. The scenario observation $\mathbf{O}(x)$ provides simulator ground truth for identifying collisions, while the runtime signals $\mathbf{M}(x)$ capture internal ADS information used to predict safety failures during operation.
For convenience, we define the \emph{scenario record} as $D(x) = (x, \mathbf{M}(x), \mathbf{O}(x))$, comprising the scenario parameters, runtime signals, and scenario observation.

\subsection{Safety-Critical Failure}\label{sec:scf}
 
ADSs are expected to satisfy various safety requirements during operation~\cite{koopman2016challenges,shalev2017rss}. Among potential safety violations, collisions involving the ego vehicle represent a critical system-level safety failure widely studied in ADS testing~\cite{av_fuzzer,kim2022drivefuzz,cheng2023behavexplor}. 
In this work, we focus on \textit{ego-vehicle collisions} as the target safety-critical failure and use the scenario observation $\mathbf{O}(x)$ as ground truth to identify their occurrence.
Formally, let $\nu(\mathbf{o}_t) \in \{0,1\}$ indicate whether an ego-vehicle collision occurs at frame $t$. 
A scenario $x$ is classified as \emph{unsafe} if $\nu(\mathbf{o}_t)=1$ for any frame $t$, and \emph{safe} otherwise.

Scenario-level labels indicate whether a collision occurs during execution, while scene-level labels characterize the safety of individual scenes. Following prior work~\cite{stocco2022thirdeye}, we define scene-level labels based on the time remaining before a collision. 
Formally, let $\tau_t$ denote the timestamp of scene $\mathbf{o}_t$. The first collision timestamp is defined as
$\tau_{\mathrm{col}}(x) = \min\{\tau_t \mid \nu(\mathbf{o}_t)=1\}$,
with $\tau_{\mathrm{col}}(x)=+\infty$ if no collision occurs.
Given a prediction horizon $H>0$, the scene label is defined as
\begin{equation}
y_t(x) =
\mathds{1}\left[
0 < \tau_{\mathrm{col}}(x)-\tau_t \leq H
\right],
\label{eq:scene-label}
\end{equation}
where $\mathds{1}[\cdot]$ denotes the indicator function.
A scene is labeled \emph{unsafe} if $y_t(x)=1$, indicating that a collision will occur within the next $H$ seconds, and \emph{safe} (i.e.,  $y_t(x)=0$) otherwise. For violation scenarios, the execution terminates upon collision, and only scenes preceding the collision are considered for scene-level labeling and evaluation.

We set $H=3$ seconds based on the advance-warning intervals used in pre-crash seat-belt pretensioning studies~\cite{lee2013precrash,zhao2017collision}. Note that this fixed horizon does not guarantee sufficient time for safety intervention.

\section{Problem Definition}
\label{sec:prob-def}

In this paper, we develop a self-evolving runtime safety monitor that predicts system-level safety failures from internal runtime signals across multiple ADS modules and improves prediction performance through automated refinement.

\begin{definition}[Safety Prediction Task]
\label{def:safety-prediction}
Given the runtime signal sequence $\mathbf{M}(x)$, the runtime signal context (RSC) at time step $t$ is defined as
\begin{equation}
    \mathbf{c}_t =
    (\mathbf{m}_{\max(1,t-L+1)}, \ldots, \mathbf{m}_t),
\end{equation}
where $L$ denotes the context window length.
A safety monitor $\mathcal{G}$ uses the current RSC $\mathbf{c}_t$ to predict whether an ego-vehicle collision will occur within the next $H$ seconds.
The risk score and predicted scene label are defined as
\begin{equation}
    s_t = \mathcal{G}(\mathbf{c}_t) \in [0,1],
    \qquad
    \hat{y}_t(x) = \mathds{1}[s_t \geq \gamma],
\end{equation}
where a higher $s_t$ indicates greater predicted collision risk and $\gamma$ denotes the decision threshold.
\end{definition}

\begin{definition}[Self-Evolving Safety Monitor]
\label{def:self-evolving-monitor}
Given an initial safety monitor $\mathcal{G}^0$ and the scenario space $\mathcal{X}$, a self-evolving monitor iteratively validates and refines its predictions by exploring additional ADS executions.
At each evolution round $r$, a set of scenarios $\mathcal{S}^r \subseteq \mathcal{X}$ is sampled based on feedback from the current monitor $\mathcal{G}^{r-1}$.
The corresponding scenario records are then used to update the monitor as
\begin{equation}
    \mathcal{G}^r =
    \mathcal{U}\left(
        \mathcal{G}^{r-1},
        \{D(x) \mid x \in \mathcal{S}^r\}
    \right),
\end{equation}
where $\mathcal{U}$ denotes the monitor refinement procedure.
After $R$ evolution rounds, we denote the final monitor by $\mathcal{G}^R$.
\end{definition}

\noindent\textbf{Overall Objective.}
Our objective is to obtain a self-evolving safety monitor that accurately identifies unsafe scenes while maintaining a low FPR, which is
\begin{equation}
\begin{aligned}
    \max_{\mathcal{G}^R,\gamma} \quad
    & \mathrm{Recall}(\mathcal{G}^R, \gamma) \\
    \mathrm{s.t.} \quad
    & \mathrm{FPR}(\mathcal{G}^R, \gamma) \leq \alpha,
\end{aligned}
\label{eq:monitor-objective}
\end{equation}
where $\mathrm{Recall}$ measures the proportion of unsafe scenes correctly identified, $\mathrm{FPR}$ measures the proportion of safe scenes incorrectly classified as unsafe, and $\alpha$ denotes the maximum allowable FPR.
To achieve this objective, we divide the problem into two connected parts:

\textit{(1) Safety Modeling.}
The first problem is to construct an initial monitor $\mathcal{G}^0$ that predicts system-level safety failures from runtime signals across multiple ADS modules. These signals reflect different aspects of the system state, making it difficult to manually identify and model their relevance to system-level safety. For example, a distant vehicle may influence the ADS's planning decisions and pose a potential collision risk, even when nearby vehicles appear safe. Therefore, the key problem is to develop a monitor that captures the relationships among heterogeneous runtime signals and the temporal evolution of system states to predict system-level safety failures.

\textit{(2) Monitor Evolution.}
Given the current monitor $\mathcal{G}^{r-1}$, the second problem is selecting informative scenarios for monitor evolution. 
Specifically, obtaining runtime signals and corresponding scenario observations requires executing the ADS in simulation, which is costly and time-consuming. Moreover, repeatedly exploring scenarios associated with similar prediction errors may concentrate the search in a few regions, leaving other monitoring weaknesses undiscovered. Therefore, the key problem is to effectively sample scenarios that expose prediction errors while balancing the exploitation of known weaknesses with the exploration of diverse scenarios.

\section{Methodology}

\begin{figure*}[!t]
    \centering
    \includegraphics[width=\textwidth]{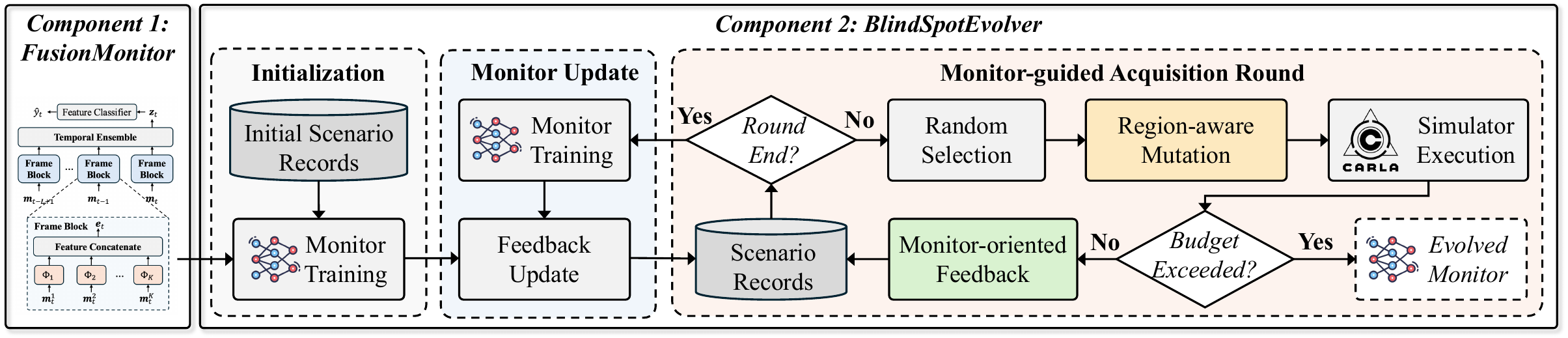}
    \caption{Overview of \tool{}.}
    \vspace{-10pt}
    \label{fig:overview}
\end{figure*}

\subsection{Overview}
\label{sec:overview}

Fig.~\ref{fig:overview} presents the overview of \tool{}, a simulation-guided framework for self-evolving runtime safety monitoring of ADSs. 
It consists of two main components, \monitor{} and \evo{}.
As shown in Fig.~\ref{fig:monitor-model}, \monitor{} models runtime signals across multiple ADS modules to predict system-level safety failures. Specifically, \monitor{} employs module-specific encoders to process heterogeneous runtime signals from different modules. It then integrates the encoded signals to represent the system state at each time step and models its temporal evolution to produce a risk score for an impending ego-vehicle collision.

\evo{} aims to continuously improve the monitor's safety prediction performance through self-evolution.
As shown in Fig.~\ref{fig:overview}, \evo{} starts with a monitor trained on an initial set of scenario records. In each evolution round, it selects scenarios likely to expose the monitor's prediction errors for simulation. The collected scenario records are then used to refine the monitor. The refined monitor guides scenario exploration in the next round.
Specifically, each evolution round combines two mechanisms. \textit{Monitor-oriented Feedback} uses the monitor's prediction errors to guide the search toward informative scenarios, while \textit{Region-aware Mutation} encourages exploration of less-explored regions to prevent the search from concentrating on similar scenarios. The selected scenarios are executed in simulation, and their records are added to the accumulated dataset to update the monitor. The updated monitor then guides the next evolution round, forming a closed loop of scenario exploration and monitor refinement. This process continues until the simulation-time budget is exhausted, returning the final evolved monitor.

\subsection{Safety Modeling: \monitor{}}
\label{sec:monitor}

\begin{figure}[!t]
    \centering
    \includegraphics[width=\linewidth]{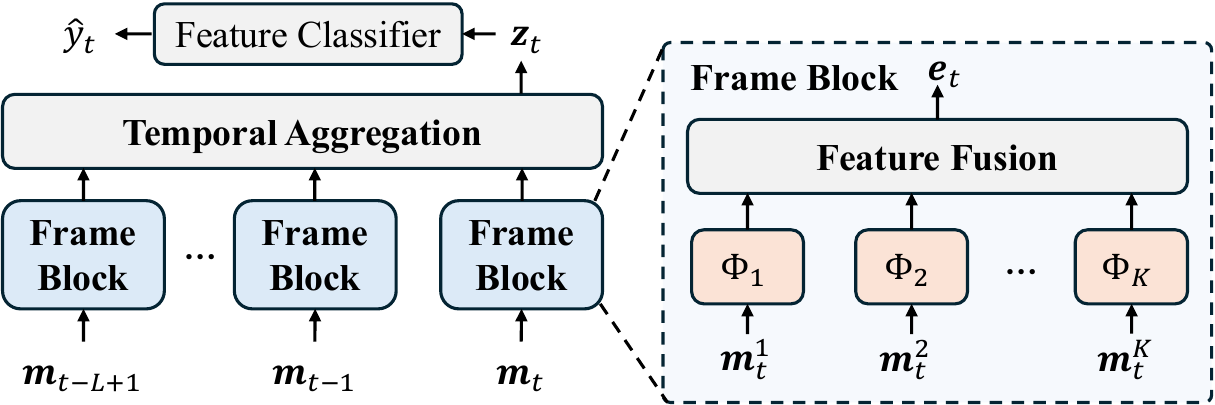}
    \caption{Architecture of \monitor{}.}
    \vspace{-10pt}
    \label{fig:monitor-model}
\end{figure}

As discussed in Section~\ref{sec:prob-def}, predicting system-level safety failures requires integrating runtime signals from multiple ADS modules. However, these signals differ substantially in structure and semantics. For example, perception produces variable-sized sets of detected objects, prediction provides their future trajectories, and planning and control generate planned paths and control commands. These differences make it difficult to obtain a unified representation of the system state and capture its temporal evolution for accurate safety prediction.

To address this challenge, we design \monitor{} with a hierarchical feature fusion architecture, as shown in Fig.~\ref{fig:monitor-model}. At each frame, the \textit{Frame Block} employs module-specific encoders to process heterogeneous runtime signals and fuses the encoded features to represent the current system state. The \textit{Temporal Aggregation} module then captures the temporal evolution of system states over the most recent \(L\) frames. Finally, a classifier produces a risk score for an ego-vehicle collision within the next \(H\) seconds. We describe these components below.

\begin{table}[!t]
    \centering
    \caption{Module-specific encoders for module-based ADSs.}
    \label{tab:module-encoders}
    \scriptsize
    \setlength{\tabcolsep}{4pt}
    \renewcommand{\arraystretch}{1.15}
    \resizebox{\linewidth}{!}{\begin{tabularx}{\columnwidth}{
        @{}
        lll
        @{}
    }
    \toprule
    \textbf{Module} &
    \textbf{Input Description} &
    \textbf{Encoder Structure} \\
    \midrule

    \textit{Perception} &
    States of detected obstacles &
    \(\{\hat{\mathbf{o}}_{t,i}\}_{i=1}^{n_t}
    \xrightarrow{\text{Shared MLP + AttnPool}}
    \mathbf{e}_t^{\mathrm{perc}}\) \\

    \midrule

    \textit{Prediction} &
    Predicted trajectories of obstacles &
    \(\{\mathbf{p}_{t,i}\}_{i=1}^{n_t}
    \xrightarrow{\text{Shared MLP + AttnPool}}
    \mathbf{e}_t^{\mathrm{pred}}\) \\

    \midrule

    \textit{Planning} &
    Planned ego trajectory &
    \(\mathbf{P}_t
    \xrightarrow{\text{Flatten + MLP}}
    \mathbf{e}_t^{\mathrm{plan}}\) \\

    \midrule

    \textit{Control} &
    Action commands &
    \(\mathbf{u}_t
    \xrightarrow{\text{MLP}}
    \mathbf{e}_t^{\mathrm{ctrl}}\) \\

    \bottomrule
    \end{tabularx}}
    \vspace{-10pt}
\end{table}

\textit{Frame Block.}
At frame $t$, the \textit{Frame Block} employs module-specific encoders to process heterogeneous runtime signals $\mathbf{m}_t = (\mathbf{m}_t^1, \ldots, \mathbf{m}_t^K)$ from multiple ADS modules.
Each encoder $\Phi_k$ transforms the runtime signals from module $k$ into a fixed-dimensional representation $\mathbf{e}_t^k = \Phi_k(\mathbf{m}_t^k) \in \mathbb{R}^{d_k}$.

Table~\ref{tab:module-encoders} summarizes these encoders using Baidu Apollo~\cite{baiduapollo} as an example.
The perception module provides detected obstacle states $\{\hat{\mathbf{o}}_{t,i}\}_{i=1}^{n_t}$, including positions, headings, and velocities.
The prediction module provides their predicted trajectories $\{\mathbf{p}_{t,i}\}_{i=1}^{n_t}$, where $\mathbf{p}_{t,i}$ represents the predicted future positions of obstacle $i$.
Here, $n_t$ denotes the number of considered obstacles at frame $t$.
To handle varying numbers of obstacles, both modules employ shared multilayer perceptrons (MLPs) to encode individual obstacles and attention pooling to aggregate their features. 
For planning, $\mathbf{P}_t$ denotes the ego vehicle's planned trajectory, represented as a sequence of future waypoints, which is flattened and encoded using an MLP.
For control, $\mathbf{u}_t$ represents the control commands, including steering, throttle, and braking, which are encoded using an MLP.
The resulting module-specific representations are then fused to characterize the system state at frame $t$.
Further implementation details are available on our website~\cite{ourweb}. The \textit{Frame Block} can be adapted to different architectures by configuring module-specific encoders according to their runtime signals.

To integrate the encoded signals across modules, we concatenate their representations and apply a frame-level MLP:
\begin{equation}
\mathbf{e}_t
=
\operatorname{MLP}_{\mathrm{frame}}
\left(
\operatorname{Concat}
\left(
\mathbf{e}_t^1,
\ldots,
\mathbf{e}_t^K
\right)
\right).
\end{equation}
The resulting representation $\mathbf{e}_t$ characterizes the system state at frame $t$ by learning nonlinear relationships among signals from different ADS modules.

\textit{Temporal Aggregation.}
To capture the temporal evolution of system states, \monitor{} aggregates the frame representations over the most recent $L$ frames.
These representations are concatenated in temporal order and processed by an MLP to learn temporal dependencies across system states.
For the first $L-1$ frames, we pad the sequence by repeating the earliest available frame representation $\mathbf{e}_1$ to maintain a fixed context length.
Specifically, we define $\mathbf{e}_i=\mathbf{e}_1$ for $i<1$.
The temporal feature is computed as
\begin{equation}
\label{eq:t-feature}
\mathbf{z}_t =
\operatorname{F}_{\mathrm{temp}}
\left(
\mathbf{e}_{t-L+1},
\ldots,
\mathbf{e}_t
\right),
\end{equation}
where $\operatorname{F}_{\mathrm{temp}}$ denotes a temporal encoder, such as a gated recurrent unit (GRU)~\cite{cho2014learning}.
The resulting temporal feature $\mathbf{z}_t$ captures how the system state evolves over the context window.

\textit{Feature Classifier.}
The temporal feature $\mathbf{z}_t$ is fed into a fully connected classifier to produce a collision-risk score
\begin{equation}
s_t =
\mathcal{G}(\mathbf{c}_t) =
\operatorname{Sigmoid}\left(
f_{\mathrm{cls}}(\mathbf{z}_t)
\right),
\end{equation}
where $f_{\mathrm{cls}}$ denotes the classifier.
The score $s_t \in [0,1]$ indicates the predicted risk of a collision within the next $H$ seconds.

\textit{Monitor Training.}
Given a set of scenario records $\mathcal{D}$, we construct a training dataset $\mathcal{T} = \{(\mathbf{c}_j, y_j)\}$, where each pair consists of a runtime signal context and its ground-truth scene label.
The contexts are extracted from runtime signals $\mathbf{M}(x)$, while the labels are derived from the corresponding scenario observations $\mathbf{O}(x)$ using Eq.~\ref{eq:scene-label}.
We train \monitor{} by:
\begin{equation}\label{eq:train}
\mathcal{G}^{*}
=
\operatorname*{arg\,min}_{\mathcal{G}}
\frac{1}{|\mathcal{T}|}
\sum_{(\mathbf{c}_j,y_j)\in\mathcal{T}}
\ell_{\mathrm{BCE}}
\left(
\mathcal{G}(\mathbf{c}_j),
y_j
\right),
\end{equation}
where $\ell_{\mathrm{BCE}}$ denotes the binary cross-entropy loss.

\subsection{Monitor Evolution: \evo{}}
\label{sec:self-evo}

The goal of \evo{} is to improve the monitor's ability to detect unsafe scenes while maintaining a low false positive rate. However, scenario acquisition faces two key challenges. First, collecting more scenario records does not necessarily improve monitoring performance. For example, a correctly predicted collision may offer limited value for refinement, whereas a missed collision or a false alarm reveals a prediction error that needs addressing. Second, repeatedly exploring scenarios near previously identified errors may concentrate the search in a narrow region, leaving other driving conditions unexplored. Therefore, effective monitor evolution requires identifying informative scenarios while maintaining diversity in scenario exploration.

To address these challenges, \evo{} combines two complementary mechanisms. \textit{Monitor-Oriented Feedback} evaluates scenario executions based on the current monitor's prediction errors, while \textit{Region-Aware Mutation} balances exploitation of monitoring weaknesses with exploration of less-sampled regions.
Fig.~\ref{fig:overview} and Algorithm~\ref{alg:tool} illustrate the evolution process. \evo{} first trains an initial monitor using the available scenario records and computes their monitoring feedback. In each iteration, it randomly selects a previously executed scenario as a seed and applies \textit{Region-Aware Mutation} to generate a new scenario. The ADS then executes this scenario in simulation, and \textit{Monitor-Oriented Feedback} evaluates the resulting scenario record. The new record and its feedback are added to the accumulated dataset. 
After collecting a fixed number of new scenario records, \evo{} updates the monitor using the accumulated data and recomputes the feedback for all records. This ensures that subsequent exploration reflects the prediction weaknesses of the updated monitor. The process continues until the budget is exhausted, returning the final evolved monitor.

\subsubsection{Feedback Calculation}
\label{sec:feedback-calculation}

To guide scenario exploration toward the current monitor's prediction weaknesses, \evo{} introduces \textit{Monitor-Oriented Feedback}. Unlike conventional safety metrics, such as minimum distance, which measure the criticality of driving scenarios, our feedback directly evaluates the monitor's prediction errors.
Formally, given an executed scenario $x$, let $\mathcal{C}(x) = (\mathbf{c}_1, \ldots, \mathbf{c}_{n_x})$ denote its runtime signal contexts, where $n_x$ is the number of evaluated scenes. At evolution round $r$, the current monitor $\mathcal{G}^r$ produces a risk score $s_i^{r} = \mathcal{G}^r(\mathbf{c}_i)$ for each context $\mathbf{c}_i$, with the corresponding ground-truth scene label $y_i(x)$. We define the scenario-level feedback as the average absolute prediction error
\begin{equation}
\label{eq:scenario-feedback}
f^r(x)
=
\frac{1}{n_x}
\sum_{i=1}^{n_x}
|s_i^{r}-y_i(x)|.
\end{equation}
A higher $f^r(x)$ indicates larger prediction errors, such as missed collisions or false alarms.

\subsubsection{Region-Aware Mutation}
\label{sec:density-aware-mutation}

In each iteration, \evo{} generates a new scenario by mutating a previously executed scenario. Repeated local mutations may concentrate the search in well-explored regions, resulting in redundant executions. Conversely, relying solely on global exploration may overlook informative scenarios near previously identified prediction errors.
To balance local exploitation and global exploration, we introduce \textit{Region-Aware Mutation}, which evaluates the exploitation potential of the region surrounding the selected seed scenario. When this potential is high, it favors local mutation to further explore nearby scenarios. Otherwise, it shifts to global exploration to discover informative scenarios in other regions. This adaptive strategy helps prevent the search from becoming concentrated in a few regions while preserving opportunities to discover informative scenarios.

\textit{Region Exploitation Potential.}
To determine which regions are worth further local exploitation, \evo{} evaluates the prediction errors and exploitation density surrounding each scenario. Given a scenario $x$, we define its neighborhood in the scenario space as
\begin{equation}
\label{eq:scenario-neighborhood}
\mathcal{N}_{\delta}(x)
=
\left\{
x' \in \mathcal{X}
\mid
d_{\mathcal{X}}(x',x) \leq \delta
\right\},
\end{equation}
where $d_{\mathcal{X}}(\cdot,\cdot)$ measures the distance between two scenarios and $\delta$ denotes the neighborhood radius. Let $\mathcal{S}_{\delta}(x)$ denote the set of distinct previously executed scenarios within this neighborhood, including $x$. 
We characterize the monitor's performance in this region using the mean and variance of the feedback scores
\begin{equation}
\label{eq:regional-feedback-statistics}
\begin{aligned}
\mu_{\delta}(x)
&=
\frac{1}{|\mathcal{S}_{\delta}(x)|}
\sum_{x_j\in\mathcal{S}_{\delta}(x)}
f^{r}(x_j),\\
\sigma_{\delta}^2(x)
&=
\frac{1}{|\mathcal{S}_{\delta}(x)|}
\sum_{x_j\in\mathcal{S}_{\delta}(x)}
\left(
f^{r}(x_j)-\mu_{\delta}(x)
\right)^2.
\end{aligned}
\end{equation}
A high $\mu_{\delta}(x)$ indicates substantial prediction errors in the region, while a high $\sigma_{\delta}^2(x)$ indicates varying monitor performance among nearby scenarios, suggesting potential blind spot boundaries. 
We further measure the local density as
\begin{equation}
\label{eq:regional-density}
D_{\delta}(x)
=
\frac{|\mathcal{S}_{\delta}(x)|-1}
{|\mathcal{S}_{\delta}(x)|-1+\kappa},
\end{equation}
where $\kappa>0$ controls how the density increases with the number of nearby executed scenarios. 
We combine these statistics to define the regional exploitation potential as
\begin{equation}
\label{eq:regional-potential}
V_{\delta}^{r}(x)
=
\frac{\mu_{\delta}(x)+2\sigma_{\delta}(x)}{2}
\left(1-D_{\delta}(x)\right).
\end{equation}
Here, $V_{\delta}^{r}(x) \in [0,1]$, with higher values indicating greater potential for local exploitation to generate informative training data that may improve the monitor.

\textit{Mutation Strategy.}
Given a seed scenario $x$, \evo{} determines whether to perform local mutation or global exploration based on its regional exploitation potential. We define the probability of local mutation as
\begin{equation}
\label{eq:normalized-regional-potential}
p_{\delta}^{r}(x)
=
p_{\min}
+
(p_{\max}-p_{\min})V_{\delta}^{r}(x),
\end{equation}
where $p_{\min}$ and $p_{\max}$ bound the probability to preserve both exploration strategies. A higher $V^{r}(x)$ favors local mutation to generate informative scenarios, while a lower value encourages global exploration. 
Consequently, given $\xi \sim \operatorname{Uniform}(0,1)$, a new scenario $x'$ is generated by
\begin{equation}
\label{eq:density-aware-mutation}
x'
\sim
\begin{cases}
\operatorname{UniformSample}
\left(\mathcal{N}_{\delta}(x)\right),
&
\text{if } \xi < p_{\delta}^{r}(x),\\[3pt]
\operatorname{UniformSample}
\left(\mathcal{X}\right),
&
\text{otherwise}.
\end{cases}
\end{equation}
Higher exploitation potential increases the probability of local mutation, encouraging \evo{} to explore nearby scenarios to find additional prediction errors. Conversely, a lower potential suggests that the monitor performs more consistently in the region, favoring global exploration to uncover monitoring weaknesses elsewhere.

\begin{algorithm}[!t]
\small
\caption{Procedure of \evo{}.}
\label{alg:tool}
\SetKwInOut{Input}{Input}
\SetKwInOut{Output}{Output}
\SetKwInOut{Para}{Parameters}
\SetKwComment{Comment}{//}{}
\SetKwFunction{Mutation}{\textbf{RegionAwareMutation}}
\SetKwFunction{Feedback}{\textbf{MonitorFeedback}}
\SetKwFunction{Train}{\textbf{MonitorTraining}}
\SetKwFunction{Finetune}{\textbf{MonitorUpdate}}
\SetKwFunction{FeedbackUpdate}{\textbf{FeedbackRefresh}}

\Input{Scenario space $\mathcal{X}$, ADS $\mathcal{A}$,
initial scenario records $\mathcal{D}^{\text{init}}$}
\Output{Evolved runtime safety monitor $\mathcal{G}^{r}$}
\Para{Evolution budget $T_{\mathrm{budget}}$, scenarios per evolution round $N_{\mathrm{round}}$
}

$r\gets 0$,
$i\gets 0$ \\

\Comment{Train the initial monitor (see Eq.~\ref{eq:train})} 
$\mathcal{G}^{r}
\gets
\Train(\text{None},\mathcal{D}^{\text{init}})$\\

\Comment{Initialize scenario feedback}
$\mathcal{D}^{r}
\gets
\FeedbackUpdate(
    \mathcal{D}^{\mathrm{init}},
    \mathcal{G}^{r}
)$ \\


\Repeat{Budget $T_{\mathrm{budget}}$ exhausted}{

    $i\gets i+1$ \\

    $x \gets \textsc{RandomSelect}(\mathcal{D}^{r})$ \\ 
    
    $x'
    \gets
    \Mutation(x, \mathcal{X},\mathcal{D}^{r})$ \\

    $\mathbf{O}(x'),\mathbf{M}(x')
    \gets
    \textsc{Simulation}(x',\mathcal{A})$ \\

    $f^r(x')
    \gets
    \Feedback(
        \mathbf{O}(x'),
        \mathbf{M}(x'),
        \mathcal{G}^{r})$ \\

    $\mathcal{D}^{r}
    \gets
    \mathcal{D}^{r}
    \cup
    \left\{
        \left(
            x',
            \mathbf{O}(x'),
            \mathbf{M}(x'),
            f^r(x')
        \right)
    \right\}$ \\

    \If{$i\bmod N_{\mathrm{round}}=0$}{
        $r\gets r+1$ \\
        
        $\mathcal{G}^{r}
        \gets
        \Train(
            \mathcal{G}^{r-1},
            \mathcal{D}^{r-1}
        )$ \\

        $\mathcal{D}^{r}
        \gets
        \FeedbackUpdate(
            \mathcal{D}^{r-1},
            \mathcal{G}^{r}
        )$ \\
        
    }
}

\Return{$\mathcal{G}^{r}$}

\end{algorithm}

\subsection{Algorithm Overview}
\label{sec:algorithm-overview}

Algorithm~\ref{alg:tool} presents the overall evolution procedure of \evo{}. Given the scenario space $\mathcal{X}$, the ADS $\mathcal{A}$, and the initial scenario records $\mathcal{D}^{\mathrm{init}}$, \evo{} alternates between scenario exploration and monitor updates to progressively improve safety prediction performance.
\evo{} first trains an initial monitor $\mathcal{G}^0$ using $\mathcal{D}^{\mathrm{init}}$ (Line~2). It then computes the monitor-oriented feedback for each initial scenario record, forming the initial record set $\mathcal{D}^0$ (Line~3). Section~\ref{sec:feedback-calculation} details the feedback calculation.
In each iteration, \evo{} randomly selects a previously executed scenario from $\mathcal{D}^r$ as the seed (Line~6). Based on the regional exploitation potential, \textit{Region-Aware Mutation} generates a new scenario through either local mutation or global sampling (Line~7). Section~\ref{sec:density-aware-mutation} details the mutation strategy. The generated scenario $x'$ is then executed with the ADS in simulation, producing runtime signals $\mathbf{M}(x')$ and the corresponding scenario observation $\mathbf{O}(x')$ (Line~8). \evo{} evaluates the execution using \textit{Monitor-Oriented Feedback} and adds the resulting scenario record with its feedback score $f^r(x')$ to $\mathcal{D}^r$ (Lines~9--10).
After every $N_{\mathrm{round}}$ new executions, \evo{} updates the monitor using the accumulated scenario records, producing new $\mathcal{G}^{r}$ (Line~13). Since the feedback scores depend on the current monitor $\mathcal{G}^{r}$, \evo{} recomputes them for all collected records using $\mathcal{G}^{r}$ (Line~14). This allows subsequent exploration to focus on the weaknesses of the updated monitor. The process repeats until the evolution budget $T_{\mathrm{budget}}$ is exhausted, returning the latest evolved monitor $\mathcal{G}^r$.

\section{Empirical Evaluation}
\label{sec:evaluation}

In this section, we empirically evaluate \tool{} by answering the following research questions:

\noindent\textbf{RQ1 (Safety Modeling Effectiveness).}
How effectively does \monitor{} detect impending safety violations compared with existing runtime safety monitoring approaches?

\noindent\textbf{RQ2 (Monitor Self-Evolution Effectiveness).}
How effectively does \evo{} identify informative scenarios to improve monitoring performance compared with alternative scenario sampling strategies under the same budget?

\noindent\textbf{RQ3 (Component Contributions).}
How do the components of \monitor{} and \evo{} affect monitoring performance and self-evolution effectiveness?

\textbf{Environment.} We conduct our experiments on Baidu \textit{Apollo}~\cite{baiduapollo}, an open-source, industrial-grade, full-stack ADS. We select \textit{Apollo} because it provides a complete, modular autonomy stack covering perception, prediction, planning, and control, and has been widely adopted in ADS research and evaluation~\cite{cheng2023behavexplor,huai2023doppelganger}. We integrate Apollo with the CARLA simulator~\cite{dosovitskiy2017carla} and enable its native sensor-processing and decision-making pipeline. Specifically, Apollo receives raw observations from simulated onboard sensors and processes them through its modules. This setup differs from prior ADS testing studies that bypass the perception stack~\cite{cheng2023behavexplor,huai2023doppelganger} by directly providing ground-truth environmental information from the simulator to the ADS. 
By retaining the complete perception-to-control pipeline, our environment preserves perception errors, thereby providing a more realistic setting for evaluating runtime monitoring of full-stack ADS executions.

\textbf{Driving Scenarios.}
We evaluate \tool{} on two safety-critical logical scenarios derived from the NHTSA pre-crash typology~\cite{najm2007pre}: \emph{S1: highway cut-in} and \emph{S2: unprotected intersection}. They represent common highway and urban driving conditions and involve multiple interacting maneuvers.
\emph{S1: highway cut-in.}
The ego vehicle interacts with six vehicles in a multi-lane highway scenario involving car following, overtaking, cut-ins from adjacent lanes, and sudden lead-vehicle braking. We expose 20 parameters controlling vehicle positions, speeds, cut-in triggers, and braking behaviors.
\emph{S2: unprotected intersection.}
The ego vehicle proceeds through a four-way intersection while interacting with five vehicles performing opposing turns, lateral crossings, and right-turn merging. We expose 10 parameters controlling their speeds and start times.

\textbf{Baselines.}
We organize the baselines into two groups corresponding to \tool{}'s two key capabilities, including runtime safety monitoring and monitor self-evolution.

\textit{(1) Runtime Safety Monitoring.}
We compare \tool{} with three representative runtime safety monitors for ADSs, including \textit{TTC} (Time-to-Collision)~\cite{ttc}, \textit{RSS} (Responsibility-Sensitive Safety)~\cite{shalev2017rss}, and \textit{SelfOracle}~\cite{stocco2020misbehaviour}. These baselines represent different monitoring approaches based on kinematic indicators, predefined safety constraints, and learning-based failure prediction. 
For \textit{TTC} and \textit{RSS}, we use vehicle states obtained from the ADS perception module rather than simulator ground truth. \textit{TTC} estimates collision risk based on the relative positions and velocities of surrounding vehicles, while \textit{RSS} evaluates longitudinal and lateral safety constraints using predefined physical models. We treat the \textit{TTC} cutoff and the longitudinal and lateral \textit{RSS} safety margins as tunable decision thresholds.
Both methods consider vehicle motion over time but rely on predefined kinematic relationships. 
\textit{SelfOracle} uses reconstruction errors and temporal smoothing to predict potential driving failures. We adapt the original implementation~\cite{stocco2020misbehaviour} to process camera sensor images.

\textit{(2) Monitor Self-Evolution.}
We evaluate the effectiveness of \evo{} against three baselines. 
\monitor{} represents the initial monitor trained without further evolution. \textit{Uniform} randomly samples scenarios from the scenario space for monitor refinement, following prior studies~\cite{torfah2022learning,stocco2022confidence}. \textit{AVFuzzer}~\cite{av_fuzzer} employs a genetic algorithm to search for scenarios that trigger ADS safety violations. The resulting scenario records are then used to refine the monitor.

\begin{table}[!t]
\centering
\caption{Dataset statistics. \#Scena.-the number of scenarios. \#Viol.-the number of collision scenarios. \#Frame-the number of frames. \#Unsafe-the number of frames labeled as unsafe.}
\label{tab:datasets}
\small
\resizebox{\linewidth}{!}{\begin{tabular}{llcccc}
\toprule
LS & Set & \#Scena. & \#Viol.\,(\%) & \#Frame & \#Unsafe\,(\%) \\
\midrule
\multirow{2}{*}{\textit{S1}}
& Train & 107 & 33\,(30.8\%) & 63,191 & 1,947\,(3.1\%) \\
& Test  & 100 & 25\,(25.0\%) & 55,147 & 1,475\,(2.7\%) \\
\midrule
\multirow{2}{*}{\textit{S2}}
& Train & 177 & 110\,(62.1\%) & 54,094 & 6,490\,(12.0\%) \\
& Test  & 177 & 92\,(52.0\%) & 62,382 & 5,428\,(8.7\%) \\
\bottomrule
\vspace{-20pt}
\end{tabular}}
\end{table}

\textbf{Datasets.}
To evaluate the safety monitors and self-evolution, we randomly sample scenarios from each logical scenario space and collect their execution data. 
Each scene is labeled as either safe or unsafe, following the definition in Section~\ref{sec:scf}.
The statistics of the collected dataset are shown in Table~\ref{tab:datasets}.

\textbf{Metrics.}
We evaluate monitoring effectiveness at both the frame and scenario levels using \textit{Precision}, \textit{Recall}, and \textit{F1-score}.
Each predicted scene label is compared with its ground-truth label.
At the scenario level, an unsafe scenario is considered successfully detected if the monitor raises at least $J$ continuous warnings before the collision within the prediction horizon. A safe scenario is falsely flagged if any warning is raised during its execution. 
Following SelfOracle~\cite{stocco2020misbehaviour}, we calibrate the decision thresholds at two target frame-level FPRs of $0.01$ and $0.05$ using safe scenarios to compare the performance of all monitors.
We report the resulting \textit{Precision}, \textit{Recall}, and \textit{F1-score}, together with \textit{AUROC} and \textit{AUPRC} to assess discrimination performance across different thresholds.

\textbf{Implementation Settings.}
We implement \tool{} using PyTorch~\cite{paszke2019pytorch}. We set the temporal context length $L$ to $10$. For monitor evolution, we set $N_{\mathrm{round}}$ to $10$ and the neighborhood radius $\delta$ to $20\%$ of each scenario parameter's range. 
The mutation probability bounds $p_{\text{min}}$ and $p_{\text{max}}$ are set to 0.15 and 0.85. All evolution methods are evaluated under the same 5 evolution rounds budget $T_{\mathrm{budget}}$.  All experiments are conducted on a Linux server equipped with an AMD EPYC 7543P CPU, an NVIDIA RTX A5000 GPU, and 256\,GB of RAM.

\subsection{RQ1: Monitoring Effectiveness}

\begin{table*}[!t]
\centering
\caption{Monitoring effectiveness on the two logical scenarios.
S1 (Highway Cut-in) and S2 (Unprotected Intersection).}
\label{tab:rq1-monitoring}
\small
\setlength{\tabcolsep}{4pt}
\renewcommand{\arraystretch}{1.05}
\resizebox{0.85\textwidth}{!}{%
\begin{tabular}{@{}clcccccccc@{}}
  \toprule
  \multirow{2.5}{*}{Scenario}
  & \multirow{2.5}{*}{Method}
  & \multicolumn{3}{c}{FPR = 0.05}
  & \multicolumn{3}{c}{FPR = 0.01}
  & \multirow{2.5}{*}{ROC$\uparrow$}
  & \multirow{2.5}{*}{PRC$\uparrow$} \\
  \cmidrule(lr){3-5}\cmidrule(lr){6-8}
  & & Prec.$\uparrow$ & Rec.$\uparrow$ & F1$\uparrow$
  & Prec.$\uparrow$ & Rec.$\uparrow$ & F1$\uparrow$
  & & \\
  \midrule
  \multirow{4}{*}{S1}
  & \textit{TTC}
  & 12.9 & 23.1 & 16.5
  & 0.0 & 0.0 & 0.0 & 71.1 & 8.4 \\
  & \textit{RSS}
  & 15.9 & 31.8 & 21.2
  & 23.5 & 10.8 & 14.8 & 68.8 & 11.2 \\
  & \textit{SelfOracle}
  & $16.0_{\pm0.4}$ & $30.6_{\pm0.9}$ & $21.0_{\pm0.5}$
  & $21.8_{\pm0.2}$ & $8.3_{\pm0.1}$ & $12.0_{\pm0.2}$
  & $78.2_{\pm0.4}$ & $12.5_{\pm0.7}$ \\
  & \monitor{}
  & $\mathbf{20.4}_{\pm2.9}$ & $\mathbf{47.0}_{\pm8.3}$ & $\mathbf{28.4}_{\pm4.3}$
  & $\mathbf{30.9}_{\pm9.4}$ & $\mathbf{17.6}_{\pm9.0}$ & $\mathbf{22.2}_{\pm9.7}$
  & $\mathbf{87.2}_{\pm1.6}$ & $\mathbf{20.8}_{\pm5.8}$ \\
  \midrule
  \multirow{4}{*}{S2}
  & \textit{TTC}
  & 32.5 & 21.6 & 25.9
  & 15.9 & 1.6 & 2.9 & 66.3 & 18.9 \\
  & \textit{RSS}
  & 37.0 & 26.9 & 31.2
  & 50.5 & 10.3 & 17.1 & 66.8 & 23.0 \\
  & \textit{SelfOracle}
  & $49.6_{\pm0.4}$ & $52.5_{\pm1.4}$ & $51.0_{\pm0.7}$
  & $67.1_{\pm1.8}$ & $23.5_{\pm2.0}$ & $34.8_{\pm2.4}$
  & $90.3_{\pm0.2}$ & $49.1_{\pm1.2}$ \\
  & \monitor{}
  & $\mathbf{56.3}_{\pm0.2}$ & $\mathbf{76.3}_{\pm3.3}$ & $\mathbf{64.8}_{\pm1.1}$
  & $\mathbf{79.4}_{\pm0.7}$ & $\mathbf{54.3}_{\pm8.5}$ & $\mathbf{64.1}_{\pm6.2}$
  & $\mathbf{96.5}_{\pm0.3}$ & $\mathbf{74.9}_{\pm2.6}$ \\
  \bottomrule
\end{tabular}%
}
\end{table*}

\begin{figure*}[!t]
    \centering
    \includegraphics[width=0.8\linewidth]{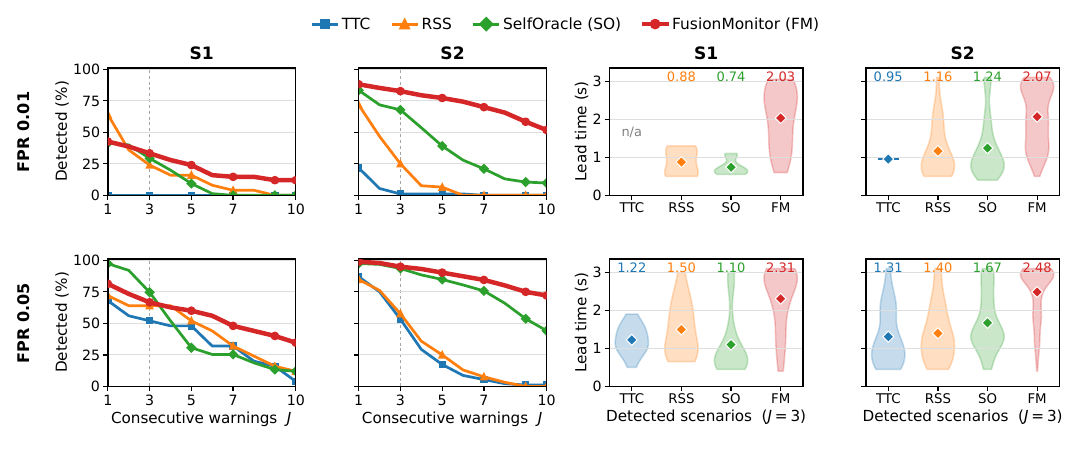}
    \caption{Unsafe scenario identification rate and distribution of warning lead times.}
    \vspace{-10pt}
    \label{fig:rq1-time}
\end{figure*}
  
In this RQ, we evaluate the effectiveness of \monitor{} against baseline runtime safety monitors. 
Learning-based monitors are independently trained three times, and their results are reported as the mean and standard deviation. \textit{TTC} and \textit{RSS} are evaluated using fixed configurations without model training. The results are presented in Table~\ref{tab:rq1-monitoring}. \textit{TTC} and \textit{RSS} are deterministic methods, so no variance is reported.

\textit{Effectiveness.}
As shown in Table~\ref{tab:rq1-monitoring}, \monitor{} achieves the best frame-level performance on both scenarios at both target FPRs. On S1, its \textit{F1-scores} reach $28.4\%$ and $22.2\%$ at target FPRs of $0.05$ and $0.01$, outperforming the strongest baseline, \textit{RSS}, by $7.2$ and $7.4$ percentage points, respectively. On S2, \monitor{} improves \textit{F1} over \textit{SelfOracle} by $13.8$ and $29.3$ percentage points. These gains are accompanied by improvements in both \textit{Precision} and \textit{Recall}, with particularly pronounced relative gains in \textit{Recall}, indicating better detection of unsafe frames under low FPRs.
The limited performance of \textit{TTC} and \textit{RSS} may stem from their reliance on predefined vehicle motion relationships and ADS-derived information. For example, in S1, safe and unsafe scenes may exhibit similar kinematic conditions before a cut-in, making impending collisions difficult to distinguish. Moreover, inaccuracies in perceived vehicle states and predicted trajectories may further affect their safety assessments. These limitations may contribute to missed warnings under strict FPR constraints, with \textit{TTC} achieving zero \textit{Recall} on S1 and only $1.6\%$ on S2 at the $0.01$ target FPR. Beyond these operating points, \monitor{} achieves the highest \textit{AUROC} and \textit{AUPRC} on both scenarios, demonstrating stronger discrimination across thresholds.

\textit{Unsafe Scenario Identification.}
Fig.~\ref{fig:rq1-time} evaluates the impact of consecutive warning requirements on unsafe scenario detection. Here, $J$ denotes the minimum number of consecutive positive predictions required to trigger a warning. As $J$ increases, detection rates generally decrease across all methods. However, \monitor{} maintains stronger detection performance under stricter requirements.
At a target FPR of $0.05$, SelfOracle initially outperforms \monitor{} on S1 at $J=1$, detecting $97.3\%$ of unsafe scenarios compared with $81.3\%$. However, at $J=10$, \monitor{} retains a detection rate of $34.7\%$, while SelfOracle drops to $12.0\%$. Similarly, on S2, \monitor{} detects $72.1\%$ of unsafe scenarios at $J=10$, compared with $44.2\%$ for SelfOracle and at most $1.1\%$ for TTC and RSS. At the stricter target FPR of $0.01$, \monitor{} achieves detection rates of $33.3\%$ and $82.6\%$ on S1 and S2 at $J=3$, exceeding SelfOracle by $4.0$ and $14.9$ percentage points, respectively. These results suggest that \monitor{} provides more temporally consistent warnings by maintaining higher detection rates as the consecutive warning requirement increases. Detailed results are available at~\cite{ourweb}.

We further evaluate how early each monitor detects impending collisions. Fig.~\ref{fig:rq1-time} shows the distribution of warning lead times for successfully detected unsafe scenarios with $J=3$. At a target FPR of $0.05$, \monitor{} achieves mean warning lead times of $2.31$ and $2.48$ seconds on S1 and S2, respectively, compared with $1.10$ and $1.67$ seconds for SelfOracle and $1.22$ and $1.31$ seconds for TTC. At the stricter target FPR of $0.01$, \monitor{} maintains mean lead times of $2.03$ and $2.07$ seconds, exceeding all three baselines shown in the figure. These results demonstrate that \monitor{} provides earlier collision warnings while maintaining strong detection performance under consecutive warning requirements.

\textit{Inference Time.}
\monitor{} incurs low inference latency, requiring only $6.26$~ms on the CPU and $10.6$~ms on the GPU to process each runtime signal context. Both are well below the ADS control cycle of $50$ to $100$~ms. The lower CPU latency is attributed to the lightweight model with only about $35$K parameters, where GPU kernel launch and synchronization overheads outweigh the benefits of parallel computation. These results demonstrate that \monitor{} supports real-time safety monitoring on the evaluated hardware without requiring dedicated GPU acceleration.

\begin{tcolorbox}[
  colback=gray!5,
  colframe=black,
  boxrule=0.9pt,
  arc=1mm,
  left=4pt,
  right=4pt,
  top=3pt,
  bottom=3pt
]
\small\textbf{Answer to RQ1:}
The designed monitor \monitor{} provides effective, reliable safety monitoring with timely warnings and millisecond-level inference overhead.
\end{tcolorbox}
\subsection{RQ2: Self-Evolution Effectiveness}

\begin{table*}[!t]
    \centering
    \caption{Effectiveness of self-evolution.}
    \label{tab:rq2-evo}
    \small
    \setlength{\tabcolsep}{5pt}
    \renewcommand{\arraystretch}{1.08}
    \resizebox{0.9\textwidth}{!}{%
    \begin{tabular}{lcccccccccccccccc}
      \toprule
      \multirow{4}{*}{Method}
      & \multicolumn{8}{c}{\textit{S1: Highway Cut-in}}
      & \multicolumn{8}{c}{\textit{S2: Unprotected Intersection}} \\
      \cmidrule(lr){2-9}\cmidrule(lr){10-17}
      & \multicolumn{3}{c}{FPR = 0.05}
      & \multicolumn{3}{c}{FPR = 0.01}
      & \multirow{2.5}{*}{ROC$\uparrow$}
      & \multirow{2.5}{*}{PRC$\uparrow$}
      & \multicolumn{3}{c}{FPR = 0.05}
      & \multicolumn{3}{c}{FPR = 0.01}
      & \multirow{2.5}{*}{ROC$\uparrow$}
      & \multirow{2.5}{*}{PRC$\uparrow$} \\
      \cmidrule(lr){2-4}
      \cmidrule(lr){5-7}
      \cmidrule(lr){10-12}
      \cmidrule(lr){13-15}
      & Prec.$\uparrow$ & Rec.$\uparrow$ & F1$\uparrow$
      & Prec.$\uparrow$ & Rec.$\uparrow$ & F1$\uparrow$
      & &
      & Prec.$\uparrow$ & Rec.$\uparrow$ & F1$\uparrow$
      & Prec.$\uparrow$ & Rec.$\uparrow$ & F1$\uparrow$
      & & \\
      \midrule

      \multicolumn{17}{l}{\textbf{Test Set}} \\
       \textit{\monitor{}*}
      & 20.4 & 47.0 & 28.4 & 30.9 & 17.6 & 22.2 & 87.2 & 20.8
      & \textbf{56.3} & 76.3 & 64.8 & 79.4 & 54.3 & 64.1 & 96.5 & 74.9 \\
      \textit{Uniform}
      & 22.5 & 51.2 & 31.3 & \textbf{37.0} & 21.6 & 27.1 & 86.2 & 24.2
      & 56.0  & 74.5 & 63.8 & \textbf{80.0} & 52.8 & 63.3 & 96.4 & 74.5 \\
      \textit{AVFuzzer}
      &  20.6 & 47.0 & 28.6 & 31.8 & 20.5 & 24.7 & 86.6 & 23.1
      & 55.8 & 76.8 & 64.6 & 79.7 & 53.8 & 64.1 & 96.4 & 74.7 \\
      \tool{}
      & \textbf{23.5} & \textbf{53.9} & \textbf{32.7} & 36.9 & \textbf{22.0} & \textbf{27.5} & \textbf{90.0} & \textbf{26.6}
      & 55.8 & \textbf{77.6} & \textbf{64.9} & 79.8 & \textbf{55.8} & \textbf{65.7} & \textbf{96.7} & \textbf{76.5} \\

      \midrule

      \multicolumn{17}{l}{\textbf{Hard Set}} \\
      \textit{\monitor{}*}
      & 67.6 & 25.3 & 36.5 & 51.9 & 1.7 & 3.2 & 85.4 & 58.1
      & 90.0 & 35.0 & 49.9 & 96.4 & 6.3 & 11.7 & 96.0 & 90.2 \\
      \textit{Uniform}
      & 73.6 & 28.5 & 40.7 & \textbf{91.3} & 5.3 & 9.9 & 83.7 & 61.4
      &  \textbf{91.2} & 41.1 & 55.6 & \textbf{99.1} & 10.2 & 18.4 & \textbf{96.2} & \textbf{91.7} \\
      \textit{AVFuzzer}
      & 69.8 & 31.1 & 42.4 & 71.5 & 10.3 & 17.3 & 84.5 & 61.5
      & 90.7 & 44.2 & 59.0 & 93.2 & 7.4 & 13.5 & 95.9 & 90.3 \\
      \tool{}
      & \textbf{76.5} & \textbf{39.8} & \textbf{52.1} & 84.8 & \textbf{13.9} & \textbf{23.3} & \textbf{89.7} & \textbf{70.4}
      & 90.0 & \textbf{47.4} & \textbf{62.0} & 93.9 & \textbf{11.5} & \textbf{20.5} & 95.9 & 90.4 \\

      \bottomrule
    \end{tabular}%
    }
\end{table*}

\begin{figure*}[!t]
    \centering
    \includegraphics[width=0.8\textwidth]{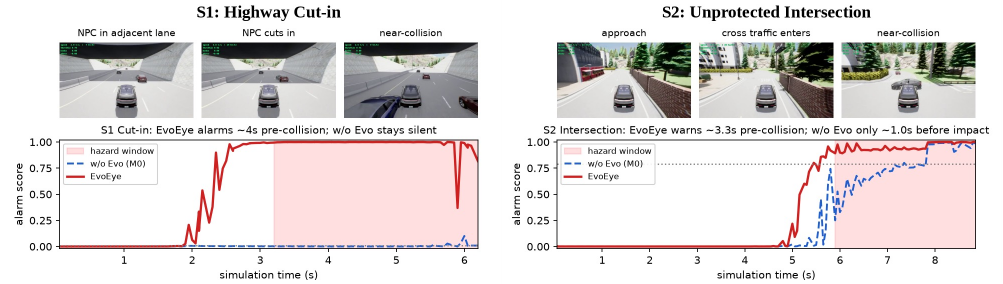}
   \caption{Illustrative cases comparing \tool{} before and after self-evolution.}
   \vspace{-10pt}
    \label{fig:case-study}
\end{figure*}

We evaluate whether \evo{} can progressively improve the performance of \monitor{} compared with two baselines, uniform sampling and AVFuzzer~\cite{av_fuzzer}.
For this comparison, we denote \monitor{}* as the monitor trained by the initial training set.
We also split a \textit{Hard Set} from the \textit{Test Set}. This subset contains unsafe scenarios that the \monitor{}* fails to detect at a target FPR of 0.01 when requiring $J=3$ consecutive warnings.
Table~\ref{tab:rq2-evo} presents the results averaged over three runs.

As shown in Table~\ref{tab:rq2-evo}, \tool{} achieves the highest \textit{F1-score} at both target FPRs, as well as the highest \textit{AUROC} and \textit{AUPRC}, on the \textit{Test Set} in both scenarios. It outperforms \textit{Uniform} and \textit{AVFuzzer} on these metrics under the same evolution budget, demonstrating the benefit of monitor-oriented scenario exploration. Compared with \monitor{}*, \tool{} improves the \textit{F1-score} from $28.4\%$ to $32.7\%$ on S1 and from $64.8\%$ to $64.9\%$ on S2 at a target FPR of $0.05$. At the stricter target FPR of $0.01$, the corresponding \textit{F1-scores} increase from $22.2\%$ to $27.5\%$ and from $64.1\%$ to $65.7\%$, respectively. These improvements are accompanied by higher \textit{Recall} across both scenarios and FPR settings, indicating that \evo{} enables \monitor{} to detect more unsafe frames under the same target FPR. The smaller gains on S2 are consistent with the stronger performance of its initial monitor.

On the \textit{Hard Set}, \tool{} substantially improves the initial monitor's \textit{F1-score} on S1 from $36.5\%$ to $52.1\%$ at a target FPR of $0.05$ and from $3.2\%$ to $23.3\%$ at $0.01$, outperforming the strongest alternative sampling method by $9.7$ and $6.0$ percentage points, respectively. On S2, \tool{} increases the \textit{F1-score} from $49.9\%$ to $62.0\%$ at a target FPR of $0.05$ and from $11.7\%$ to $20.5\%$ at $0.01$, outperforming the strongest alternative sampling method by $3.0$ and $2.1$ percentage points, respectively.
These results demonstrate that \evo{} improves the ability 
of \monitor{} to detect unsafe frames in challenging 
scenarios while maintaining the FPRs. 

\textit{Case Analysis.}
Fig.~\ref{fig:case-study} illustrates how \evo{} improves collision warnings in two representative cases. The red and blue curves represent the risk scores of \tool{} and the initial \monitor{} without evolution (\textit{w/o Evo}), respectively. 
In S1, as an NPC vehicle cuts into the ego lane, \tool{} triggers a warning approximately $4$ seconds before the collision, while \textit{w/o Evo} remains below the warning threshold and fails to detect the impending collision.
In S2, as cross traffic enters the intersection, \tool{} raises a warning approximately $3.3$ seconds before the collision and maintains a high risk score. In contrast, \textit{w/o Evo} triggers a warning only about $1$ second before the collision, leaving substantially less time for a response.
These cases demonstrate that \evo{} effectively improves \monitor{} by reducing missed collision warnings and enabling earlier detection of impending collisions.

\begin{tcolorbox}[
  colback=gray!5,
  colframe=black,
  boxrule=0.9pt,
  arc=1mm,
  left=4pt,
  right=4pt,
  top=3pt,
  bottom=3pt
]
\small\textbf{Answer to RQ2.}
\evo{} improves overall monitoring performance and the detection of previously missed unsafe frames, outperforming uniform and violation-oriented scenario sampling.
\end{tcolorbox}

\subsection{RQ3: Component Contributions}

\begin{table}[!t]
\centering
\caption{Ablation results of different ADS module inputs.}
\label{tab:rq3-monitor-ablation}
\footnotesize
\setlength{\tabcolsep}{2.5pt}
\renewcommand{\arraystretch}{1.1}
\resizebox{0.9\linewidth}{!}{%
\begin{tabular}{@{}c|l|cccc@{}}
\toprule
\textbf{Scenario} & \textbf{Variant}
& \textbf{F1@.01}$\uparrow$
& \textbf{F1@.05}$\uparrow$
& \textbf{ROC}$\uparrow$
& \textbf{PRC}$\uparrow$ \\
\midrule

\multirow{7}{*}{S1}
& Perc. only
& $26.1_{\pm13.2}$ & $25.7_{\pm4.9}$
& $82.6_{\pm1.5}$ & $22.8_{\pm9.3}$ \\

& Pred. only
& $\mathbf{26.6}_{\pm10.5}$ & $26.9_{\pm2.9}$
& $78.6_{\pm3.0}$ & $\mathbf{23.7}_{\pm9.5}$ \\

& Plan. only
& $6.0_{\pm1.6}$ & $19.2_{\pm2.6}$
& $84.2_{\pm1.6}$ & $12.3_{\pm1.9}$ \\

& Ctrl. only
& $4.3_{\pm2.3}$ & $11.5_{\pm3.9}$
& $79.7_{\pm3.3}$ & $8.7_{\pm2.4}$ \\

& Perc.+Pred.
& $26.1_{\pm1.1}$ & $27.9_{\pm2.7}$
& $83.1_{\pm2.8}$ & $21.1_{\pm2.7}$ \\

& Plan.+Ctrl.
& $5.1_{\pm0.8}$ & $19.8_{\pm0.7}$
& $85.4_{\pm0.6}$ & $11.5_{\pm0.3}$ \\

& \monitor{}
& $22.2_{\pm9.7}$ & $\mathbf{28.4}_{\pm4.3}$
& $\mathbf{87.2}_{\pm1.6}$ & $20.8_{\pm5.8}$ \\

\midrule

\multirow{7}{*}{S2}
& Perc. only
& $64.8_{\pm6.6}$ & $63.2_{\pm0.9}$
& $95.7_{\pm0.4}$ & $70.9_{\pm2.2}$ \\

& Pred. only
& $\mathbf{66.0}_{\pm2.9}$ & $63.4_{\pm0.5}$
& $96.0_{\pm0.4}$ & $72.5_{\pm2.2}$ \\

& Plan. only
& $36.3_{\pm0.4}$ & $52.0_{\pm0.1}$
& $93.7_{\pm0.1}$ & $53.7_{\pm0.6}$ \\

& Ctrl. only
& $33.7_{\pm2.0}$ & $51.3_{\pm1.3}$
& $93.2_{\pm0.3}$ & $50.7_{\pm1.0}$ \\

& Perc.+Pred.
& $61.2_{\pm5.2}$ & $62.9_{\pm0.8}$
& $95.9_{\pm0.3}$ & $71.1_{\pm1.3}$ \\

& Plan.+Ctrl.
& $34.2_{\pm0.7}$ & $52.0_{\pm0.9}$
& $93.6_{\pm0.2}$ & $53.8_{\pm1.3}$ \\

& \monitor{}
& $64.1_{\pm6.2}$ & $\mathbf{64.8}_{\pm1.1}$
& $\mathbf{96.5}_{\pm0.3}$ & $\mathbf{74.9}_{\pm2.6}$ \\

\bottomrule
\end{tabular}%
}
\end{table}

\begin{table}[!t]
\centering
\caption{Impact of temporal context length $L$.}
\label{tab:rq3-temporal-ablation}
\footnotesize
\setlength{\tabcolsep}{6pt}
\renewcommand{\arraystretch}{1.15}
\resizebox{0.8\linewidth}{!}{
\begin{tabular}{@{}clcccc@{}}
\toprule
\textbf{Scenario} & \textbf{Metric}
& $L=1$ & $L=5$ & $\boldsymbol{L=10}$ & $L=20$ \\
\midrule
\multirow{2}{*}{S1}
& AUROC $\uparrow$
& 84.4 & \textbf{88.8} & 87.2 & 85.9 \\
& AUPRC $\uparrow$
& 18.3 & 16.3 & \textbf{20.8} & 15.9 \\
\midrule
\multirow{2}{*}{S2}
& AUROC $\uparrow$
& 95.0 & 95.2 & \textbf{96.5} & {96.4} \\
& AUPRC $\uparrow$
& 64.5 & 68.5 & 74.9 & \textbf{75.7} \\
\bottomrule
\end{tabular}
}
\vspace{-10pt}
\end{table}

\subsubsection{Monitor Ablation}

Table~\ref{tab:rq3-monitor-ablation} presents the ablation results of \monitor{} with different ADS module inputs. We consider six configurations, including four using individual modules, two combining either upstream (perception and prediction) or downstream (planning and control) modules.

Across both scenarios, upstream variants achieve higher \textit{F1 scores} and \textit{AUPRC} than downstream variants, suggesting that much of the information relevant to collision risks is captured by the perceived traffic scene and its predicted evolution. Nevertheless, downstream signals also provide valuable risk information, as evidenced by the \textit{AUROC} of $85.4\%$ on S1 and $93.6\%$ on S2 achieved by the planning-and-control variant, potentially reflecting inadequate ADS responses to hazardous situations. By integrating both types of signals, \monitor{} achieves the highest mean scores in five of the eight scenario-metric combinations, including the highest \textit{F1 scores} at the $0.05$ target FPR and the highest \textit{AUROC} on both scenarios. Although the prediction-only variant performs better on certain metrics, these results suggest that jointly monitoring the traffic scene and the ADS's response to it provides complementary information for effective safety prediction.

We further analyze the impact of temporal context length $L$. As shown in Table~\ref{tab:rq3-temporal-ablation}, increasing $L$ from $1$ to $10$ improves AUPRC by $2.5$ and $10.4$ percentage points on S1 and S2, respectively. However, further increasing $L$ to $20$ degrades both AUROC and AUPRC on S1, while providing only a marginal AUPRC improvement on S2. These results suggest that the benefits of longer temporal contexts are scenario dependent and do not consistently increase with $L$. We therefore select $L=10$ as a balanced setting that maintains strong monitoring performance across both scenarios.


\begin{table}[!t]
\centering
\caption{Ablation results of \evo{}.}
\label{tab:rq3-evolution-ablation}
\footnotesize
\setlength{\tabcolsep}{3pt}
\renewcommand{\arraystretch}{1.05}
\resizebox{0.6\linewidth}{!}{\begin{tabular}{@{}lcccc@{}}
\toprule

\multirow{2.5}*{Variant} & \multicolumn{2}{c}{\textit{S1}} & \multicolumn{2}{c}{\textit{S2}} \\
\cmidrule(lr){2-3}
\cmidrule(lr){4-5}

& ROC$\uparrow$
& PRC$\uparrow$ 
& ROC$\uparrow$
& PRC$\uparrow$ \\
\midrule
\textit{w/o Monitor}
& 89.2 & 23.4 & 96.4 & 75.2 \\
\textit{w/o Density}
& 87.5	& 25.4 & 96.5 & 76.3 \\
\evo{} & \textbf{90.0} & \textbf{26.6} & \textbf{96.7} & \textbf{76.5}  \\

\bottomrule
\end{tabular}}
\vspace{-10pt}
\end{table}

\subsubsection{Self-Evolution}

Table~\ref{tab:rq3-evolution-ablation} evaluates the contributions of the monitor score and density information to \evo{}. Specifically, \textit{w/o Monitor} removes the monitor score from the feedback, while \textit{w/o Density} removes density information. Both variants retain the mutation mechanism, which uses the available feedback to guide scenario exploration.
As shown in Table~\ref{tab:rq3-evolution-ablation}, removing the monitor score reduces \textit{AUROC} and \textit{AUPRC} by $0.8$ and $3.2$ percentage points on S1, and by $0.3$ and $1.3$ points on S2, respectively. Removing density information decreases \textit{AUROC} and \textit{AUPRC} by $2.5$ and $1.2$ points on S1, and by $0.2$ and $0.2$ points on S2, respectively. These results suggest that both feedback components contribute to the effectiveness of \evo{}, with their combination having the best monitoring performance in terms of \textit{AUROC} and \textit{AUPRC} across both scenarios.

\begin{tcolorbox}[
  colback=gray!5,
  colframe=black,
  boxrule=0.9pt,
  arc=1mm,
  left=4pt,
  right=4pt,
  top=3pt,
  bottom=3pt
]
\small\textbf{Answer to RQ3:}
Cross-module and temporal modeling improve safety monitoring performance, while effective self-evolution depends on the sampling design.
\end{tcolorbox}

\subsection{Threats to Validity}
\label{sec:threats}

\textit{(1) Internal Validity.}
The randomness involved in model training may affect monitoring performance. To mitigate this threat, we independently train learning-based monitors three times and report the mean and standard deviation of the results.
Monitoring performance may also depend on the choice of driving scenarios. To reduce this threat, we evaluate the monitors in two distinct scenarios involving different driving conditions and collision patterns.

\textit{(2) External Validity.}
Our evaluation focuses on Apollo, which may limit the generalizability of our findings to other ADSs. To mitigate this threat, our framework design is not tied to Apollo and can accommodate different ADS architectures by adapting its input signals and replacing the corresponding encoder blocks. Nevertheless, its effectiveness on other ADSs requires further validation.
The use of simulation may also limit the applicability of our findings to real driving environments. To reduce this threat, we retain Apollo's complete sensing and control pipeline in CARLA rather than providing simulator ground truth directly to the ADS. However, further evaluation in real driving environments is needed.

\section{Discussion}
\label{sec:discussion}

This section discusses the practical applicability of \tool{}.

\textbf{Sim-to-Real Gap.}
Simulation cannot fully reproduce real-world sensor behavior, vehicle dynamics, and traffic interactions~\cite{hu2024simulation,lambertenghi2025reality}, which may limit the transferability of learned monitors to real-world driving. Although \monitor{} relies on internal ADS runtime signals, differences in their distributions may still affect prediction accuracy. Recent advances in high-fidelity simulation, such as NVIDIA's ReSim~\cite{yang2025resim}, offer more realistic driving scenarios and may help narrow this gap. Nevertheless, further validation using real-world ADS execution data is needed to assess the transferability of \tool{}.

\textbf{Monitor Prediction Horizon.}
We adopt a fixed prediction horizon of $H=3$ seconds, following prior studies on pre-crash seat-belt pretensioning that consider warning intervals of 0.5 to 3 seconds~\cite{stocco2022thirdeye,lee2013precrash,zhao2017collision}. This horizon enables consistent evaluation of monitoring performance but does not guarantee sufficient time for safety intervention. The required intervention time varies with vehicle speed, braking capability, and road conditions. Longer prediction horizons may provide earlier warnings but also introduce greater prediction uncertainty. Future work will investigate prediction horizons that account for different driving conditions and intervention requirements.

\textbf{Monitor Self-Evolution.}
In this work, self-evolution refers to automated monitor refinement through offline simulation rather than online adaptation during deployment. The evolved monitor is deployed with fixed model parameters for runtime safety prediction. Although our implementation focuses on a learning-based collision monitor, the feedback-guided evolution process could also support monitors based on explicit safety specifications~\cite{mehdipour2023formal,sun2024redriver}. This would require mechanisms to revise and validate specifications based on discovered prediction errors. Enabling safe online adaptation remains future work.

\textbf{Application to AI Systems.}
The underlying approach of \tool{} may extend to other modular AI systems, such as robotic systems. Such extensions require a system-level failure oracle, observable runtime signals, and an executable scenario space. Module-specific encoders can be adapted to different signal structures, while feedback-guided evolution can support iterative monitor refinement. Further evaluation is needed to validate its effectiveness beyond ADSs.

\section{Related Work}
\label{sec:related-work}

\textbf{Safety Monitoring for ADSs.}
Existing runtime safety monitoring approaches for ADSs employ different techniques to assess potential safety failures. Specification-based approaches evaluate vehicle behaviors against predefined safety constraints~\cite{mehdipour2023formal}. TTC~\cite{ttc} estimates collision time based on relative vehicle motion, while RSS~\cite{shalev2017rss} defines safe distances and driving responses. Reachability analysis further examines possible future vehicle states~\cite{althoff2014online}. Other approaches, such as REDriver~\cite{sun2024redriver}, use signal temporal logic to monitor planned trajectories. These methods provide interpretable safety assessments but depend on predefined models and safety properties.
Model-focused approaches instead assess potential failures using information from specific driving models. SelfOracle~\cite{stocco2020misbehaviour} uses input reconstruction errors, while ThirdEye~\cite{stocco2022thirdeye} leverages neural attention maps. Farid et al.~\cite{farid2023task} further assess trajectory prediction errors based on their impact on downstream planning. These approaches primarily focus on specific models or failure mechanisms. In this paper, \tool{} integrates heterogeneous runtime signals across multiple ADS modules and models the temporal evolution of system states for safety monitoring.

\textbf{Scenario Search for ADSs.}
The vast ADS scenario space makes exhaustive testing impractical, motivating automated scenario search for discovering safety violations~\cite{cheng2023behavexplor,han2021preliminary,av_fuzzer}. Existing techniques include guided fuzzing~\cite{cheng2023behavexplor,av_fuzzer,pang2022mdpfuzz}, evolutionary search~\cite{gambi2019automatically,tang2021collision,zhou2023specification}, metamorphic testing~\cite{han2020metamorphic}, surrogate model guided search~\cite{haq2022efficient,zhong2022neural}, and reinforcement learning~\cite{haq2023many,feng2023dense,lu2022learning}. These approaches primarily aim to identify ADS safety violations or explore diverse driving behaviors. However, scenarios that reveal safety violations may provide limited value for monitor refinement when the monitor can already predict these failures correctly.
Recent studies have also investigated improving runtime monitors using additional execution data. Stocco and Tonella~\cite{stocco2022confidence} employ confidence driven retraining to adapt failure predictors, while Torfah et al.~\cite{torfah2022learning} investigate monitor learning through counterexample guided refinement. These studies demonstrate the benefits of monitor adaptation but do not jointly address targeted scenario exploration for monitor improvement. In this paper, \tool{} uses the monitor's prediction errors to guide scenario search and acquire informative scenarios for monitor refinement.
\section{Conclusion}

In this paper, we present \tool{}, a simulation-guided self-evolving safety monitoring approach for ADSs. 
\tool{} integrates two key components: \monitor{}, which learns from heterogeneous and temporally dependent runtime signals to predict impending collisions, and \evo{}, which uses \textit{Monitor-Oriented Feedback} and \textit{Region-Aware Mutation} to iteratively acquire informative scenario records and improve the monitor within a limited budget. 
Our evaluation shows that \tool{} provides effective, timely collision warnings, incurs low runtime overhead, and progressively improves its monitoring capability through self-evolution. 


\section{Data Availability}
\noindent Our artifact is available at~\cite{ourweb}.

\section*{Acknowledgment}

\noindent 
This research is partially supported by the Ministry of Education, Singapore under its Academic Research Fund Tier 2 (Proposal ID: T2EP20223-0043) and the Lee Kong Chian Fellowship. Any opinions, findings, conclusions, or recommendations expressed in this material are those of the author(s) and do not reflect the views of the Ministry of Education, Singapore. Lionel C. Briand’s contribution was partially funded by the Canada Discovery Grant program of the Natural Sciences and Engineering Research Council of Canada (NSERC) as well as the Research Ireland grant 13/RC/209.

\bibliographystyle{IEEEtran}
\bibliography{reference}

\end{document}